\documentclass[12pt,a4paper]{article}

\usepackage[latin1]{inputenc}
\usepackage{amsmath}
\usepackage{amssymb}
\usepackage{amsfonts}
\usepackage{graphicx}
\usepackage{color}
\usepackage{bm}
\usepackage[latin1]{inputenc}
\usepackage[T1]{fontenc}

\usepackage{lscape}

\usepackage{latexsym}
\usepackage{amssymb}
\usepackage{epsfig}

\usepackage{rotating}

\newtheorem{theorem}{Theorem}
\newtheorem{corollary}[theorem]{Corollary}

\newtheorem{remark}[theorem]{Remark}

\newcommand{\bib}{\par\noindent\hangindent=0.5 true cm\hangafter=1}

\def\i{{\rm i}}

\begin{document}

\title{\bf
 A model specification test for the variance function in nonparametric regression
}

\author{
Juan Carlos Pardo-Fern\'andez$^1$
\footnote{\textbf{Corresponding author.} Departamento de Estat\'istica e I.O., Facultade de Ciencias Econ\'{o}micas e Empresariais, Campus Universitario, 36310 Vigo, Spain. e-mail: juancp@uvigo.es. Telephone: +34 986 813505. Fax: +34 986 812 401.} ,
 M. Dolores Jim\'enez-Gamero$^2$
\footnote{Departamento de Estad\'istica e I.O., Facultad de Matem\'{a}ticas, Calle Tarfia s.n., 41012 Sevilla, Spain. e-mail: dolores@us.es.}
\vspace{10pt}\\
\small{$^1$Departamento de Estat\'istica e I.O., Universidade de Vigo, Spain}\\
\small{$^2$Departamento de Estad\'istica e I.O., Universidad de Sevilla, Spain}\\
}

\date{}

\maketitle

\begin{abstract}
The problem of testing for the parametric form of the conditional variance is considered in a fully nonparametric regression model.
A test statistic based on a weighted $L_2$-distance between the empirical characteristic functions of residuals constructed under the null hypothesis and under the alternative is proposed and studied theoretically. The null asymptotic distribution of the test statistic is obtained and employed to approximate the critical values. Finite sample properties of the proposed test are numerically investigated in several Monte Carlo experiments. The developed results assume independent data. Their extension to dependent observations is also discussed.
\end{abstract}

\bigskip


\section{Introduction}

Let $(X,Y)$ be a bivariate random vector which is assumed to follow the nonparametric regression model
\begin{equation} \label{model}
 Y=m(X)+\sigma(X)\varepsilon,
\end{equation}
where $ m(x) = E(Y \mid X=x) $ is the regression function,  $\sigma^2(x) = Var(Y \mid X=x)$ is the conditional variance function and $\varepsilon$ is the regression error, which is assumed to be independent of $X$. Note that, by construction, $E(\varepsilon)=0$ and $Var(\varepsilon)$=1. The covariate $X$ is one-dimensional and continuous with probability density function (pdf)  $f$ and compact support $R$. The pdf $f$ is assumed to be positive on $R$. The regression function, the variance function, the distribution of the error and the distribution of the covariate are unknown and no parametric models are assumed for them.

Parametric specifications for either the regression function or the variance function are quite attractive among practitioners since they describe the relation between the response and the covariate in a concise way. Because of this reason, there are a number of papers dealing with the parametric modelling of the conditional mean and the conditional variance of $Y$, given $X$. This paper proposes a new goodness-of-fit test for the parametric form of the conditional variance. Specifically,
on the basis of independent observations 
from \eqref{model}, we wish to test the null hypothesis
\[
H_0 : \sigma^2(x)= \sigma^2(x; \theta), \quad \mbox{for some }\theta \in \Theta \subset \mathbb{R}^p,  \quad \forall x \in R,
\]
where $\sigma^2(\cdot; \theta)$ represents a parametric model for the conditional variance function, against the general alternative
\[
H_1 : H_0 \text{ is not true}.
\]

Several tests for $H_0$ have been proposed in the specialised literature. Some of them were designed for testing homoscedasticity (see, for example, Liero, 2003,
Dette and Marchlewski, 2010); others assume that the regression function has a known parametric  form (see, for example,  Koul and Song, 2010, Samarakoon and Song, 2011, 2012);
others were built for fixed design points (see, for example, Dette and Hetzler, 2009a,b);  others  detect contiguous alternatives converging to the null at a rate slower that $n^{-1/2}$ (see, for example, Wang and Zhou, 2007, Samarakoon and Song, 2011, 2012). Although initially the test in Koul and Song (2010) was designed for parametric regression functions, they also provide the theory for  a version  where the mean function is nonparametrically estimated.

The test in Dette et al. (2007) (henceforth DNV) does not possess any of the above cited cons. The methodology that will be proposed in the present paper is, in a certain sense, close to that in DNV. Next, we describe our proposal and the precise meaning of such closeness.

Let
\[\varepsilon=\{Y-m(X)\}/\sigma(X)\]
be the regression error and let
\[\varepsilon_0=\{Y-m(X)\}/\sigma(X;\theta)\]
be the  error under the null hypothesis, which will be called null error.
 In Lemma 1 of DNV it is shown that, under the assumption that all the moments of $\varepsilon$ exist, $H_0$ is true if and only if $\varepsilon$ and $\varepsilon_0$ have the same distribution. The existence of infinite moments can be relaxed to the continuity of $\sigma(\cdot)$ and $0<E(\varepsilon^4)<\infty$, as shown for a similar problem in Theorem 1-(i) of Pardo-Fern\'andez et al. (2015b). The  equivalence between $H_0$ and the equality of distributions of $\varepsilon$ and $\varepsilon_0$ can be interpreted in terms of any function characterizing a distribution such as the cumulative distribution function (cdf) or the characteristic function (cf). DNV used this characterization of the null hypothesis and  constructed a test for $H_0$ based on comparing estimators of the cdfs of $\varepsilon$ and $\varepsilon_0$. Alternatively, in the current paper we will compare estimators of the cfs of $\varepsilon$ and $\varepsilon_0$. As it will be seen, this approach presents some advantages, both theoretical and practical. This approach is in the line of other testing procedures based on the ecf, such as those in Hu\v{s}kov\'{a} and Meintanis (2009, 2010) and Pardo-Fern\'{a}ndez et al. (2015a,b).

The paper is organized as follows. Section~\ref{theteststatistic} introduces the test statistic $T_n$. Section~\ref{Main results} studies some large sample properties of $T_n$, namely,  its asymptotic null distribution, as well as its behavior against fixed and contiguous alternatives. It is shown that the test that rejects the null hypothesis for large values of $nT_n$ is able to detect any fixed alternative; it is also shown that such a test  can detect local alternatives converging to the null at the rate $n^{-1/2}$, $n$ denoting the sample size. These properties are shared with the test in  DNV, but the one proposed in this paper requires weaker assumptions. Specifically, we do not need to assume that the regression errors have a smooth bounded pdf.

To determine what a large value of $nT_n$ means, it is needed to know its null distribution or at least a consistent null distribution estimator. Although the bootstrap is easy to implement and, under certain conditions, yields consistent null distribution estimators, it may be very time-consuming. Section~\ref{sect.prac} gives a consistent null distribution estimator that is based on estimating the asymptotic null distribution. Its calculation is really fast.

The finite sample performance of the proposed procedure was numerically assessed by some simulation experiments. A summary of the obtained results is reported in Section~\ref{Numerical results}. Finally, Section~\ref{Discussion&extensions} outlines some possible extensions of the test studied in this paper. All proofs are sketched in the Appendix.

The following notation will be used along the paper:
all vectors are column vectors; if $x \in \mathbb{R}^k$, $x^T$ denotes its transpose 
and $\|x\|$ denotes its Euclidean norm; all limits in this paper are taken when   $n \rightarrow \infty$; 
$\stackrel{\mathcal{L}}{\rightarrow}$ denotes convergence in distribution;
$\stackrel{P}{\rightarrow}$ denotes convergence in probability;
for any complex number $z=a+\i b$, $Re(z)=a$ is its real part, $Im(z)=b$ is its imaginary part, $\i=\sqrt{-1}$ and 
$|z|=\sqrt{a^2+b^2}$ is its modulus;
an unspecified integral denotes integration over the whole real line $\mathbb{R}$;
for a given non-negative real-valued function $w$  and for any complex-valued measurable function $g$, we denote $\|g\|_w=\left(\int  |g(t)|^2 w(t)dt \right)^{1/2}$ to the norm in the Hilbert space $L_2(w)=\{g:\mathbb{R}\to \mathbb{C}:\, \|g\|_w<\infty\}$.

\section{The test statistic} \label{theteststatistic}

Let $(X_1,Y_1)^T,  \ldots, (X_n,Y_n)^T$ be independent copies  from $(X,Y)^T$, which follows model \eqref{model}. The associated errors, $\varepsilon_1=\{Y_1-m(X_1)\}/\sigma(X_1), \ldots, \varepsilon_n=\{Y_n-m(X_n)\}/\sigma(X_n)$,
as well as the null errors associated with the sample,
$\varepsilon_{01}=\{Y_1-m(X_1)\}/\sigma(X_1;\theta), \ldots,
\varepsilon_{0n}=\{Y_n-m(X_n)\}/\sigma(X_n;\theta)$,
 are not observable because $m$, $\sigma$ and $\theta$ are unknown, so they must be replaced by adequate estimators.
 With this aim we use nonparametric estimators for $m$ and $\sigma^2$  based on kernel smoothing techniques. Let $K$ denote a nonnegative kernel function defined in $\mathbb{R}$, let $0<h_n\equiv h \rightarrow 0$ be the bandwidth or smoothing parameter and $K_h(x)=h^{-1}K(x/h)$. We use the following estimators for the functions $m$ and  $\sigma^2$:
\begin{equation}\label{m&s}
\hat{m}(x)=\sum_{j=1}^{n} W_{j}(x)Y_{j}, \quad \hat{\sigma}^2(x)=\sum_{j=1}^{n} W_{j}(x)\{ Y_{j}-\hat{m}(x)\}^2,
\end{equation}
where  the quantities $\{W_{j}(x)\}_{j=1}^n$ are either local-linear or  Nadaraya-Watson weights (see, for example, Fan and Gijbels, 1996, pages 20 and 15, respectively). Under the model assumptions that will be stated in the next section, the results in this article are valid for both kinds of weights. As for the parameter estimator, $\hat{\theta}$, we take, as in DNV,
 \[
 \hat{\theta}=\underset{\theta \in \Theta}{\arg \min} \, S_n(\theta),
 \] 
 where
 \begin{equation} \label{ese_ene}
 S_n(\theta)=\frac{1}{n}\sum_{j=1}^n [\{Y_j-\hat{m}(X_j)\}^2-\sigma^2(X_j; \theta)]^2.
 \end{equation}
After replacing all unknowns by the above described estimators, we have two sets of resi\-duals: the (ordinary) residuals
\[
 \hat{\varepsilon}_{j} = \frac{Y_{j}-\hat{m}(X_{j})}{\hat{\sigma}(X_{j})}, \quad j=1,\ldots,n,
\]
and the null residuals
\[
 \hat{\varepsilon}_{0j} = \frac{Y_{j}-\hat{m}(X_{j})}{{\sigma}(X_{j};\hat{\theta})}, \quad j=1,\ldots,n.
 \]
 Let $\hat{\varphi}$ and $\hat{\varphi}_0$ denote the empirical characteristic functions (ecf) associated with these two sets of residuals, that is,
 \[
 \hat{\varphi}(t)=\frac{1}{n}\sum_{j=1}^n \exp (\i t \hat{\varepsilon}_{j}),\qquad  \text{and}  \qquad
 \hat{\varphi}_0(t)=\frac{1}{n}\sum_{j=1}^n \exp (\i t \hat{\varepsilon}_{0j}).
 \]
 Under the null hypothesis $\hat{\varphi}$ and $ \hat{\varphi}_0$ should be close, but not under alternatives. To measure the distance between cfs we will employ an $L_2(w)$-norm, for some non-negative function $w$ so that $\int w(t)dt<\infty$, that is,
 \[
 T_n=\|\hat{\varphi}-\hat{\varphi}_0 \|_w^2,
 \]
which converges in probability to (see Theorem \ref{limite} below)
\begin{equation} \label{tau}
\tau=\|{\varphi}-{\varphi}_0 \|_w^2,
\end{equation}
 where ${\varphi}$ and ${\varphi}_0$ stand for the cfs of $\varepsilon$ and $\varepsilon_0$, respectively, that is, ${\varphi}(t)=E(e^{\i t \varepsilon})$ and ${\varphi}_0(t)=E(e^{\i t \varepsilon_0})$.
Observe that $\tau\geq 0$, and that under $H_0$, $\tau$ vanishes.  As a consequence, under $H_0$, $T_{n}$ should be ``very small''. It will be seen that if $w>0$, then $\tau=0$ if and only if $H_0$ is true. We then conclude that, any value of $T_{n}$ which is ``significantly large'' should lead to the rejection of $H_0$. In practice, given a significance level, a threshold value above which $H_0$ is rejected needs to  be established. To this end we need to study the null distribution of $T_{n}$. Since this distribution is unknown, the asymptotic null distribution will be employed as an approximation. This will be done in Section 4.

\begin{remark}\label{practical-calculation}
From Lemma 1 in  Alba-Fern\'andez et al.
 (2008), an alternative expression for $T_{n}$, which is useful from a computational point of view, is given by
$$
n^2T_{n}=  \sum_{j,s=1}^{n}I_w(\hat{\varepsilon}_{j}-\hat{\varepsilon}_{s}) +
\sum_{j,s=1}^{n}I_w(\hat{\varepsilon}_{0j}-\hat{\varepsilon}_{0s})-
2\sum_{j,s=1}^{n}I_w(\hat{\varepsilon}_{j}-\hat{\varepsilon}_{0s}),
$$
where $ I_w(t)=\int \cos (tx)w(x)dx$.
\end{remark}

\begin{remark}
 The empirical ecfs $\hat{\varphi}(t)$ and  $\hat{\varphi}_0(t)$ are periodic functions, so if we  consider
as test statistic
\[ \int |\hat{\varphi}(t)-\hat{\varphi}_0(t)|^2dt,\]
then its value would be $\infty$. To avoid this undesirable fact, a weight function $w$, non-negative and  with finite integral $\int w(t)dt<\infty$, is introduced in the definition of the test statistic in order to ensure its finiteness
\[ \int |\hat{\varphi}(t)-\hat{\varphi}_0(t)|^2w(t)dt<\infty.\]
In addition, since $\int w(t)dt<\infty$, the presence of the weight function is useful to downweight the persistent oscillations  of the ecfs in the tails (it is well-known that the ecf is a good estimator of the population cf in compact intervals containing 0, but its behaviour in the tails is not so good). With this aim, one can choose $w$ as the indicator function of any compact interval containing 0, which gives weight zero to the tails. Nevertheless, in such a case the resulting test will not be consistent against all fixed alternatives. Because of the above considerations, an advisable choice for $w$ is a pdf positive on the whole real line giving high probability to intervals centered at 0, such as the normal law, which is our choice for $w$ in the numerical experiments.
\end{remark}

\section{Main results} \label{Main results}

In order to study the limit behaviour of the test statistic $T_{n}$, we first need to introduce some assumptions.

\begin{description}

\item[{\bf Assumption A:}]

\item[\qquad A.1]  (i) $X$ has a compact support $R$. (ii) $f$, $m$ and $\sigma$ are two times continuously differentiable on $R$. (iii) $\inf_{x\in R} f(x)>0$ and $\inf_{x\in R} \sigma(x)>0$. (iv) $E(\varepsilon^4)<\infty$.

\item[\qquad A.2]$K$ is a twice continuously differentiable symmetric pdf with compact support.

\item[\qquad A.3] $nh^4\rightarrow 0$  and  $nh^2/\ln n \rightarrow \infty$.
\end{description}

These assumptions are mainly needed to guarantee the uniform consistency of the kernel estimators  $\hat{m}$ and $\hat{\sigma}$. Moreover, all results in this Section keep on being true if different bandwidths, say $h_1$ and $h_2$, are taken for estimating the conditional mean  and the conditional variance, whenever they meet A.3. This becomes evident from the proofs. Nevertheless, in order to simplify notation, we take $h_1=h_2=h$. Unlike the methods based on the empirical cdf, observe that we do not impose any restriction on  the distribution of the errors, like the existence of a pdf. So the results in this paper could be used when the distribution of the errors is arbitrary. An example with errors having a mixed-type distribution is provided in Section \ref{Numerical results}.

Next we give conditions on the parametric function $\sigma(x;\theta)$ in the null hypothesis to ensure the consistency and asymptotic normality of $\hat{\theta}$, which estimates $\theta_0$, where
\[
\theta_0=\underset{\theta \in \Theta}{\arg \min}\, S(\theta)
\]
 and
 \begin{equation} \label{ese}
 S(\theta)=E\left([\{Y-m(X)\}^2-\sigma^2(X; \theta)]^2\right).
\end{equation}
Throughout this paper we assume that $\theta_0$ exists and is unique. Let 
\[
\dot\sigma^2(x;\theta)=\frac{\partial}{ \partial \theta}\sigma^2(x;\theta)
\]
and
\begin{equation} \label{Omega}
\Omega=E\{\dot \sigma^2(X;\theta_0)\dot \sigma^2(X;\theta_0)^T\}.
\end{equation}

\begin{description}

\item[{\bf Assumption B:}]

\item[\qquad B.1]  For any $\delta>0$, $\displaystyle \inf_{\|\theta-\theta_0\|>\delta} S(\theta)-S(\theta_0)>0$.  

\item[\qquad B.2] $\Theta$ is compact and $\theta_0$ belongs to the interior of $\Theta$.

\item[\qquad B.3] (i) $\sigma^2(x;\theta)$ is    twice continuously differentiable with respect to $\theta$ and the derivatives are continuous in $(x,\theta)$, for all $x$ and $\theta $. (ii) $\inf_{x\in R, \, \theta\in \Theta_0} \sigma(x;\theta)>0$, for some $\Theta_0 \subseteq \Theta$ such that $\theta_0 \in int \Theta_0$.

\item[\qquad B.4] The matrix $\Omega$  defined in \eqref{Omega} is non-singular.
\end{description}

Under Assumptions A and B, routine calculations show that $\hat{\theta}$ is a consistent estimator for $\theta_0$,
\begin{equation} \label{thetanullprobability}
\hat{\theta} \stackrel{P}{\rightarrow} \theta_0.
\end{equation}
In addition, under the null hypothesis,
\begin{equation} \label{thetanull}
\sqrt{n}(\hat{\theta}-{\theta}_0)=\Omega^{-1}\frac{1}{\sqrt{n}}\sum_{j=1}^n(\varepsilon_j^2-1)\sigma^2(X_j;\theta_0)\dot \sigma^2(X_j;\theta_0)+o_P(1).
\end{equation}
A sketch of  proofs for \eqref{thetanullprobability} and \eqref{thetanull} is included in the Appendix.

\subsection{Asymptotic null distribution} \label{AsymptoticNullDistribution}

We first give a result that provides an asymptotic approximation for $\sqrt{n}\{\hat{\varphi}(t)-\hat{\varphi}_0(t)\}$ when the null hypothesis is true.
Let
\[
\varphi'(t)=
\frac{\partial}{\partial t}Re\varphi(t)+\i
\frac{\partial}{\partial t}Im\varphi(t)=\i E\left[\varepsilon \exp(\i t\varepsilon)\right],
\]
which exists because $E(|\varepsilon|)<\infty$, and let
\[\mu=E\{\dot \sigma^2(X;\theta_0)/\sigma^2(X;\theta_0)\}.\]

\begin{theorem} \label{null-expansion}
Suppose that Assumptions A and B hold.
If $H_0$ is true, then
\[\sqrt{n}\{\hat{\varphi}(t)-\hat{\varphi}_0(t)\}=-\frac{t}{2} \varphi'(t)\frac{1}{\sqrt{n}}\sum_{j=1}^n(\varepsilon^2_j-1)\{1-\mu^T\Omega^{-1}\dot \sigma^2(X_j;\theta_0)\sigma^2(X_j;\theta_0)\}+tR_1(t)+t^2R_2(t),\]
with $\sup_t |R_k(t)|=o_P(1)$, $k=1,2$.
\end{theorem}

\begin{corollary} \label{null-distribution}
Suppose that the assumptions in Theorem \ref{null-expansion} hold and that the weight function satisfies
\begin{equation}  \label{condicionw}
w(t) \geq 0, \text{ for all } t \in \mathbb{R}, \text{ and } \int t^4 w(t)dt<\infty,
\end{equation} then
\[nT_n  \stackrel{\mathcal{L}}{\longrightarrow}  \upsilon^2Z^2,\]
where $Z$ is a standard normal variate and $\upsilon^2=\upsilon^2_1 \upsilon^2_2 \upsilon^2_3$, with $\upsilon^2_1=\frac{1}{4} \| t \varphi'(t) \|_w^2$, $\upsilon^2_2=E\{(\varepsilon^2-1)^2\}$ and $\upsilon^2_3=E\left[\{1-\mu^T\Omega^{-1}\dot \sigma^2(X;\theta_0)\sigma^2(X;\theta_0)\}^2\right]$.
\end{corollary}

\begin{remark} \label{sigma0}
To define the null residuals, DNV estimated $\sigma^2(x;\theta)$ by means of
\[
 \hat{\sigma}_0^2(x;\hat{\theta})=\sum_{j=1}^{n} W_{j}(x) \sigma^2(X_{j};\hat{\theta}),
 \]
which is a smoothed version of the null parametric estimator of the variance function constructed with the same weights used for the nonparametric estimator $\hat{\sigma}^2$. The results in Theorem \ref{null-expansion} and Corollary \ref{null-distribution} keep on being true if $\hat{\sigma}_0^2(x;\hat{\theta})$ is used  instead of $\sigma^2(x;\hat{\theta})$, and if it is also assumed that $\sigma(x;\theta)$ is continuously differentiable with respect to $x$ and the derivative is continuous in $(x,\theta)$, for all $x$ and $\theta $.
\end{remark}

\begin{remark} \label{homoscedasticity}
An   instance of the null hypothesis $H_0$ is that of testing for  homoscedasticity, that is,  $H_0: \sigma^2(x;\theta)=\theta$, for some $\theta>0$.
 Straightforward calculations show that the main term of the
asymptotic representation in Theorem  \ref{null-expansion} vanishes  and, as a
 consequence, the limit distribution in Corollary \ref{null-distribution} degenerates to a
Dirac measure that is concentrated at 0. The test in DNV also shares this peculiarity.
\end{remark}

\begin{remark}
The observation in Remark \ref{homoscedasticity} persists if other estimators of $\theta$ are  used. For example,   a natural way of estimating $\theta$ for the considered test statistic could be
 $\tilde{\theta}=\underset{\theta \in \Theta}{\arg \min} \, T_n $. Routine calculations show that under the assumptions in Theorem \ref{limite}, $\tilde{\theta}$ converges in pro\-ba\-bility to $\theta_1=\underset{\theta \in \Theta}{\arg \min} \,\tau $, where $\tau$ is as defined in \eqref{tau}. In addition, under the assumptions in Corollary \ref{null-distribution}, it can be derived that when using $\tilde{\theta}$ instead of $\hat{\theta}$,
\begin{equation} \label{TMD}
n T_n  \stackrel{\mathcal{L}}{\longrightarrow} \nu^2 Z^2,
\end{equation}
where $Z$ is a standard normal variate and $\nu^2=\nu^2_1 \nu^2_2 \nu^2_3$, with $\nu_1^2=\upsilon^2_1$, $\nu_2^2=\upsilon^2_2$, $\nu^2_3=\{1-\mu^T \Omega_1^{-1}\mu\}^2$, and
$\Omega_1=E\{\dot \sigma^2(X;\theta_1)\dot \sigma^2(X;\theta_1)^T/\sigma^4(X;\theta_1)\}$. Again straightforward calculations show that the limit distribution in \eqref{TMD} degenerates to a
Dirac measure that is concentrated at the point 0 when $H_0: \sigma^2(x;\theta)=\theta$ holds, for some $\theta>0$.
\end{remark}

\subsection{Consistency}

This subsection studies the limit behavior of the test statistic $nT_n$ under fixed alternatives. With this aim, we first derive the limit in probability of $T_n$.

\begin{theorem} \label{limite}
Suppose that Assumptions A, B.1 and  B.3 (ii) hold and $w$ satisfies \eqref{condicionw}. In addition, suppose that  $\sigma^2(x;\theta)$ is   continuously differentiable with respect to $\theta$ and the derivative is continuous in $(x,\theta)$, for all $x$ and $\theta $. Then $T_n=\tau+o_P(1)$, where $\tau$ is as defined in \eqref{tau}.
\end{theorem}

Let $\alpha\in (0,1)$ be arbitrary but fixed. As an immediate consequence of Theorem \ref{limite} and  Corollary \ref{null-distribution}, the test that rejects $H_0$ when $nT_n\geq t_{\alpha}$, where $t_{\alpha}$  is the $1-\alpha$ percentile of the null distribution of $nT_n$ or any consistent estimator of it, is consistent against all fixed alternatives such that $\tau>0$, in the sense that $P(nT_n\geq t_{\alpha})\to 1$. Since  two distinct characteristic functions can be equal in a finite interval (see, for example, Feller, 1971; p. 479), to ensure that $\tau=0$ if and only if $H_0$ is true, it suffices to take $w$ such that $w(t)>0$ for all $t \in \mathbb{R}$.

\subsection{Local power}
This subsection studies the limit behavior of the test statistic $nT_n$ under the local alternative
\[
H_{1,n}:\,  \sigma^2(x)= \sigma^2(x; \theta_0)+n^{-1/2}r(x), \quad \forall x \in R,
\]
for some $\theta_0 \in \Theta \subset \mathbb{R}^p$ and
some function $r$ which will be assumed to satisfy the next assumption.  

\begin{description}

\item[{\bf Assumption L:}] $r$ is two times continuously differentiable on $R$.
\end{description}

\begin{theorem} \label{cont-expansion}
Suppose that Assumptions A, B and L hold. If $H_{1,n}$ is true, then
\begin{eqnarray*}
\sqrt{n}\{\hat{\varphi}(t)-\hat{\varphi}_0(t)\} & = &-\frac{t}{2} \varphi'(t)\left[\frac{1}{\sqrt{n}}\sum_{j=1}^n(\varepsilon^2_j-1)\{1-\mu^T\Omega^{-1}\dot \sigma^2(X_j;\theta_0)\sigma^2(X_j;\theta_0)\}+d\right] \\
 & & +tR_1(t)+t^2R_2(t),
\end{eqnarray*}
with
\[
d=E\{r(X)D(X;\theta_0)/\sigma^2(X;\theta_0)\}, \quad D(X;\theta_0)=1-\sigma^2(X;\theta_0)\mu^T\Omega^{-1}\dot\sigma^2(X;\theta_0)
\]
and $\sup_t |R_k(t)|=o_P(1)$, $k=1,2$.
\end{theorem}

\begin{corollary} \label{cont-distribution}
Suppose that the assumptions in Theorem \ref{cont-expansion} hold and that the weight function satisfies \eqref{condicionw},
then
\[
nT_n  \stackrel{\mathcal{L}}{\longrightarrow} { \upsilon_1^2(\upsilon_2\upsilon_3Z+d)^2},
\]
where $\upsilon_1$, $\upsilon_2$, $\upsilon_3$ and $Z$ are as defined in Corollary  \ref{null-distribution}.
\end{corollary}

Comparing Theorem 2 in DNV and Corollary \ref{cont-distribution} above, it catches our attention that the tests in both papers have a similar behaviour under local alternatives, as the distribution is the one under the null hypothesis plus a drift. This drift, denoted by $d$, coincides for both types of statistics, the one based on the empirical distribution function and the one based on the empirical characteristic function. Therefore, the tests in DNV and the one in the current paper will have non-trivial power against the same local alternatives.

An anonymous referee asked us to comment about contiguous alternatives for which $ d\neq 0$  or  $d=0$. It is not easy to say something general about such cases, since the value of $d$ depends on the expression of $\sigma^2(x;\theta_0)$, on $\theta_0$ and on the distribution of $X$, and there are many possibilities for all of them. In addition, $\theta_0$ and  the distribution of $X$ are unknown in applications. Nevertheless, it is clear that both cases, $d\neq 0$ and $d = 0$, may exist. On the one hand, whenever $D(x;\theta_0)$ is not identically equal to 0 for all $x\in R$, one can always find a function $r$ so that the associated value of $d$ is not equal to $0$ by simply taking $r(x)=D(x;\theta_0)$. On the other hand, for each $\sigma^2(x;\theta_0)$, each $\theta_0$ and each distribution of $X$, one can always find a  function $r$ so that the associated value of $d$ is equal to $0$ by taking
\[
r(x)=\frac{\sigma^2(x;\theta_0)}{D(x;\theta_0)}-\frac{1}{E\{D(X;\theta_0)/\sigma^2(X;\theta_0)\}},
\]
if $D(x;\theta_0)\neq 0$, and arbitrarily defined for those $x\in R$ such that $D(x;\theta_0)=0$.

\section{Estimation of the null distribution}\label{sect.prac}

The result in Corollary \ref{null-distribution}  gives the  asymptotic null distribution of the proposed test statistic, which is unknown because it depends on unknown quantities.  Therefore, the asymptotic null distribution cannot be directly used to approximate the null distribution of $T_n$. Two solutions can be considered: either (a) to approximate the null distribution by a resampling procedure, such as the bootstrap, or (b) to construct an approximation of the asymptotic null distribution. The first approach was also considered in DNV, who  employed a bootstrap procedure based on smoothed residuals. The same bootstrap procedure could be used to approximate the null distribution of $T_n$ in the current setting.

For the second  possibility, recall that the  asymptotic  null distribution is $\upsilon^2 Z^2$, which is unknown because  $\upsilon^2$  is unknown.  Each term in the expression of $\upsilon^2$ can be consistently estimated as follows:
\[
\hat{\upsilon}_1^2=\frac{1}{4} \| t \hat{\varphi}'(t) \|_w^2=\frac{-1}{4n^2}\sum_{j,k=1}^n\hat{\varepsilon}_j \hat{\varepsilon}_kI''_w(\hat{\varepsilon}_j-\hat{\varepsilon}_k),
\]
where $I''_w(t)=\frac{\partial^2}{\partial t^2}I_w(t)$ and $I_w$ is as defined in Remark \ref{practical-calculation},
\[
\hat{\upsilon}_2^2=\frac{1}{n}\sum_{j}^n(\hat{\varepsilon}_j^2-1)^2
\]
and
\[
\hat{\upsilon}_3^2=\frac{1}{n}\sum_{j=1}^n\left\{1-\hat{\mu}^T \hat{\Omega}^{-1}\dot \sigma^2(X_j;\hat{\theta})\sigma^2(X_j;\hat{\theta})\right\}^2,
\]
with $\hat{\mu}=\frac{1}{n}\sum_{j=1}^n \dot \sigma^2(X_j;\hat{\theta})/\sigma^2(X_j;\hat{\theta})$, $\hat{\Omega}=\frac{1}{n}\sum_{j=1}^n \dot \sigma^2(X_j;\hat{\theta})\dot \sigma^2(X_j;\hat{\theta})^T$.

\section{Numerical results} \label{Numerical results}

Simulation experiments were conducted in order to study the finite sample performance of the null distribution approximation discussed in Section \ref{sect.prac} as well as to compare the power of the proposed test to that of the Cram\'{e}r-von Mises-type statistic studied in DNV, which is based on bootstrap. Although easy to implement, the bootstrap is very time-consuming, specially for large sample sizes. In contrast, we will use the asymptotic null distribution of $T_n$ approximated as explained in the previous section.

Recall that the procedure in DNV employs a smooth bootstrap of residuals to approximate the null distribution of the test statistic. This means that the bootstrap samples are constructed with bootstrap residuals of the form $\tilde{\varepsilon}^\ast=\varepsilon^\ast+\nu N$, where $\varepsilon^\ast$ is randomly drawn from the empirical distribution of the sample of (standarized) residuals $\{\hat{\varepsilon}_i, i=1,\ldots,n\}$, $N$ is an independently drawn mean-zero random variable and the value $\nu$ controls the smoothing. Recently, Neumeyer and Van Keilegom (2017) proved that the smoothed ($\nu>0$) and the non-smoothed ($\nu=0$) versions of this kind of bootstrap are asymptotically equivalent. Unreported simulations showed that a better level approximation for the test in DNV is obtained for the non-smoothed case, therefore only the case $\nu=0$ will be shown in the tables below.

We have considered the following two scenarios:
\begin{description}
\item[Scenario 1.] The null hypothesis is
\[
H_0: \sigma^2(x;\theta)=1+\theta x^2, \quad  \mbox{for some }\theta>0, \quad \forall x \in [0,1],
\]
and the true conditional variance function is of the form
\[
 \sigma^2(x)=1+3x^2+2.5c\sin(2\pi x).
\]

\item[Scenario 2.] The null hypothesis is
\[
H_0: \sigma^2(x;\theta)=\exp(\theta x), \quad  \mbox{for some }\theta>0, \quad \forall x \in [0,1],
\]
and the true conditional variance function is of the form
\[
 \sigma^2(x)=\exp(x)+c \sin(2\pi x).
\]
\end{description}
In both cases, three values of $c$ are considered: $0$, $0.5$ and $1$. The null hypothesis holds when $c=0$, whereas $c=0.5$ and $c=1$ yield models under the alternative hypothesis. In both scenarios the regression function is  $m(x)=1+\sin(2\pi x)$ and the covariate is uniformly distributed in $[0,1]$, that is, $R=[0,1]$. The first scenario was also considered in DNV.

Nonparametric estimation of the regression and variance  functions was performed by means of local-linear estimators. Notice that when local-linear weights are used, the estimator of the conditional variance is not guaranteed to be positive. In order to be able to run the simulations, in the unlikely event of obtaining a negative value of the variance, we have replaced it by its true value. The kernel function is the kernel of Epanechnikov $K(u)=0.75 (1-u^2)I(-1<u<1)$ and the smoothing parameters for $m$ and $\sigma^2$ are chosen separately by least-squares cross-validation.  Concerning the choice of the weight function $w$, we follow Pardo-Fern\'andez et al. (2015a,b) as well as the references therein, and take the pdf of a standard normal distribution. All tables display the observed proportion of rejections in 1000 simulated data sets and the corresponding 95\%-confidence intervals for the true proportion of rejections as an indication of the simulation error.

Tables~\ref{table1} and \ref{table2} display the results under scenarios 1 and 2, respectively, when the distribution of the regression error is a standard normal and the sample sizes are $100$ and $200$. The asymptotic null distribution of $T_n$ produces an overestimation of the level when $n=100$, but the approximation is correct when $n=200$ (note that the 95\%-confidence intervals do not cover the corresponding nominal levels for $n=100$, but they do for $n=200$, except for one case in table~\ref{table2}). The bootstrap-based statistic of DNV yields a level approximation which is closer to the nominal value for both sample sizes. In terms of power, the new test based on characteristic functions yields better results, specially under scenario 2. We should advise that a fair power comparison only makes sense for $n=200$, as for $n=100$ the level obtained with $T_n$ is overestimated, specially under scenario 2. We have also tried the bootstrap approximation for $T_n$ and the results were very close to the ones obtained with the asymptotic distribution.

\begin{table}
\caption{Observed proportion of rejections in 1000 simulated data sets under scenario~1 (between brackets, 95\%-confidence interval for the true proportion of rejections). The distribution of the errors is standard normal. \medskip \label{table1}}
\centering
\begin{tabular}{cccccccccc}
\hline
& & & \multicolumn{3}{c}{$T_n$}  & & \multicolumn{3}{c}{DNV}\\
\cline{4-6}  \cline{8-10}
$c$ & $n$ & $\alpha:$ & 0.100 & 0.050 & 0.025 & & 0.100 & 0.050 & 0.025\\
\hline
\vspace{-0.25cm} 0 & 100 &  & 0.118 & 0.068 & 0.041 &  & 0.095 & 0.049 & 0.022\\
 &  &  & \scriptsize{$(0.098,0.138)$} & \scriptsize{$(0.052,0.084)$} & \scriptsize{$(0.029,0.053)$} &  & \scriptsize{$(0.077,0.113)$} & \scriptsize{$(0.036,0.062)$} & \scriptsize{$(0.013,0.031)$}\\
\vspace{-0.25cm}  & 200 &  & 0.095 & 0.047 & 0.025 &  & 0.105 & 0.051 & 0.028\\
 &  &  & \scriptsize{$(0.077,0.113)$} & \scriptsize{$(0.034,0.060)$} & \scriptsize{$(0.015,0.035)$} &  & \scriptsize{$(0.086,0.124)$} & \scriptsize{$(0.037,0.065)$} & \scriptsize{$(0.018,0.038)$}\\
\vspace{-0.25cm} 0.5 & 100 &  & 0.676 & 0.558 & 0.473 &  & 0.435 & 0.314 & 0.224\\
 &  &  & \scriptsize{$(0.647,0.705)$} & \scriptsize{$(0.527,0.589)$} & \scriptsize{$(0.442,0.504)$} &  & \scriptsize{$(0.404,0.466)$} & \scriptsize{$(0.285,0.343)$} & \scriptsize{$(0.198,0.250)$}\\
\vspace{-0.25cm}  & 200 &  & 0.910 & 0.849 & 0.779 &  & 0.813 & 0.706 & 0.592\\
 &  &  & \scriptsize{$(0.892,0.928)$} & \scriptsize{$(0.827,0.871)$} & \scriptsize{$(0.753,0.805)$} &  & \scriptsize{$(0.789,0.837)$} & \scriptsize{$(0.678,0.734)$} & \scriptsize{$(0.562,0.622)$}\\
\vspace{-0.25cm} 1 & 100 &  & 0.984 & 0.961 & 0.924 &  & 0.883 & 0.770 & 0.652\\
 &  &  & \scriptsize{$(0.976,0.992)$} & \scriptsize{$(0.949,0.973)$} & \scriptsize{$(0.908,0.940)$} &  & \scriptsize{$(0.863,0.903)$} & \scriptsize{$(0.744,0.796)$} & \scriptsize{$(0.622,0.682)$}\\
\vspace{-0.25cm}  & 200 &  & 0.997 & 0.995 & 0.991 &  & 0.996 & 0.983 & 0.965\\
 &  &  & \scriptsize{$(0.994,1.000)$} & \scriptsize{$(0.991,0.999)$} & \scriptsize{$(0.985,0.997)$} &  & \scriptsize{$(0.992,1.000)$} & \scriptsize{$(0.975,0.991)$} & \scriptsize{$(0.954,0.976)$}\\
\hline
\end{tabular}
\end{table}

\begin{table}
\caption{Observed proportion of rejections in 1000 simulated data sets under scenario~2 (between brackets, 95\%-confidence interval for the true proportion of rejections). The distribution of the errors is standard normal.\medskip \label{table2}}
\centering
\begin{tabular}{cccccccccc}
\hline
& & & \multicolumn{3}{c}{$T_n$}  & & \multicolumn{3}{c}{DNV}\\
\cline{4-6}  \cline{8-10}
$c$ & $n$ & $\alpha:$ & 0.100 & 0.050 & 0.025 & & 0.100 & 0.050 & 0.025\\
\hline
\vspace{-0.25cm} 0 & 100 &  & 0.157 & 0.094 & 0.060 &  & 0.106 & 0.059 & 0.033\\
 &  &  & \scriptsize{$(0.134,0.180)$} & \scriptsize{$(0.076,0.112)$} & \scriptsize{$(0.045,0.075)$} &  & \scriptsize{$(0.087,0.125)$} & \scriptsize{$(0.044,0.074)$} & \scriptsize{$(0.022,0.044)$}\\
\vspace{-0.25cm}  & 200 &  & 0.121 & 0.065 & 0.030 &  & 0.118 & 0.053 & 0.021\\
 &  &  & \scriptsize{$(0.101,0.141)$} & \scriptsize{$(0.050,0.080)$} & \scriptsize{$(0.019,0.041)$} &  & \scriptsize{$(0.098,0.138)$} & \scriptsize{$(0.039,0.067)$} & \scriptsize{$(0.012,0.030)$}\\
\vspace{-0.25cm} 0.5 & 100 &  & 0.280 & 0.178 & 0.110 &  & 0.148 & 0.065 & 0.036\\
 &  &  & \scriptsize{$(0.252,0.308)$} & \scriptsize{$(0.154,0.202)$} & \scriptsize{$(0.091,0.129)$} &  & \scriptsize{$(0.126,0.170)$} & \scriptsize{$(0.050,0.080)$} & \scriptsize{$(0.024,0.048)$}\\
\vspace{-0.25cm}  & 200 &  & 0.393 & 0.286 & 0.199 &  & 0.309 & 0.180 & 0.121\\
 &  &  & \scriptsize{$(0.363,0.423)$} & \scriptsize{$(0.258,0.314)$} & \scriptsize{$(0.174,0.224)$} &  & \scriptsize{$(0.280,0.338)$} & \scriptsize{$(0.156,0.204)$} & \scriptsize{$(0.101,0.141)$}\\
\vspace{-0.25cm} 1 & 100 &  & 0.570 & 0.454 & 0.356 &  & 0.341 & 0.210 & 0.132\\
 &  &  & \scriptsize{$(0.539,0.601)$} & \scriptsize{$(0.423,0.485)$} & \scriptsize{$(0.326,0.386)$} &  & \scriptsize{$(0.312,0.370)$} & \scriptsize{$(0.185,0.235)$} & \scriptsize{$(0.111,0.153)$}\\
\vspace{-0.25cm}  & 200 &  & 0.830 & 0.751 & 0.647 &  & 0.712 & 0.569 & 0.416\\
 &  &  & \scriptsize{$(0.807,0.853)$} & \scriptsize{$(0.724,0.778)$} & \scriptsize{$(0.617,0.677)$} &  & \scriptsize{$(0.684,0.740)$} & \scriptsize{$(0.538,0.600)$} & \scriptsize{$(0.385,0.447)$}\\
\hline
\end{tabular}
\end{table}

As mentioned in Section~\ref{Main results}, the proposed method does not require the existence of a pdf for the regression error. The distribution of $\varepsilon$ can be arbitrary, as long as model \eqref{model} holds. To check the practical performance of the test for non-continuous errors, we have tried a mixed-type distribution. More precisely, $\varepsilon$ is $0$ with probability $0.5$ and $N(0,\sqrt{2})$ with probability 0.5. The results under scenarios 1 and 2 are displayed in Table~\ref{table3}. Larger sample sizes are needed to achieve a good level approximation: for $n=200$ the level is still overestimated, whereas for $n=400$ the approximation is correct. This is also shown by the 95\%-confidence intervals: they cover the corresponding nominal levels for $n=400$, but not for $n=200$. Notice that the method in DNV requires the existence of a pdf for $\varepsilon$ and thus cannot be applied to this setting.

\begin{table}
\caption{Observed proportion of rejections in 1000 simulated data sets under scenarios 1 and 2 when the statistic $T_n$ is employed (between brackets, 95\%-confidence interval for the true proportion of rejections). The distribution of the errors is of a mixed-type.\medskip \label{table3}}\centering
\begin{tabular}{cccccccccc}
\hline
& & & \multicolumn{3}{c}{Scenario 1}  & & \multicolumn{3}{c}{Scenario 2}\\
\cline{4-6}  \cline{8-10}
$c$ & $n$ & $\alpha:$ & 0.100 & 0.050 & 0.025 & & 0.100 & 0.050 & 0.025\\
\hline
\vspace{-0.25cm} 0 & 200 &  & 0.146 & 0.072 & 0.033 &  & 0.175 & 0.104 & 0.061\\
 &  &  & \scriptsize{$(0.124,0.168)$} & \scriptsize{$(0.056,0.088)$} & \scriptsize{$(0.022,0.044)$} &  & \scriptsize{$(0.151,0.199)$} & \scriptsize{$(0.085,0.123)$} & \scriptsize{$(0.046,0.076)$}\\
\vspace{-0.25cm} & 400 &  & 0.095 & 0.049 & 0.023 &  & 0.110 & 0.057 & 0.028\\
 &  &  & \scriptsize{$(0.077,0.113)$} & \scriptsize{$(0.036,0.062)$} & \scriptsize{$(0.014,0.032)$} &  & \scriptsize{$(0.091,0.129)$} & \scriptsize{$(0.043,0.071)$} & \scriptsize{$(0.018,0.038)$}\\
\vspace{-0.25cm}0.5 & 200 &  & 0.692 & 0.570 & 0.456 &  & 0.312 & 0.211 & 0.129\\
 &  &  & \scriptsize{$(0.663,0.721)$} & \scriptsize{$(0.539,0.601)$} & \scriptsize{$(0.425,0.487)$} &  & \scriptsize{$(0.283,0.341)$} & \scriptsize{$(0.186,0.236)$} & \scriptsize{$(0.108,0.150)$}\\
\vspace{-0.25cm} & 400 &  & 0.884 & 0.801 & 0.722 &  & 0.408 & 0.271 & 0.189\\
 &  &  & \scriptsize{$(0.864,0.904)$} & \scriptsize{$(0.776,0.826)$} & \scriptsize{$(0.694,0.750)$} &  & \scriptsize{$(0.378,0.438)$} & \scriptsize{$(0.243,0.299)$} & \scriptsize{$(0.165,0.213)$}\\
\vspace{-0.25cm}1 & 200 &  & 0.974 & 0.949 & 0.912 &  & 0.621 & 0.490 & 0.388\\
 &  &  & \scriptsize{$(0.964,0.984)$} & \scriptsize{$(0.935,0.963)$} & \scriptsize{$(0.894,0.930)$} &  & \scriptsize{$(0.591,0.651)$} & \scriptsize{$(0.459,0.521)$} & \scriptsize{$(0.358,0.418)$}\\
\vspace{-0.25cm} & 400 &  & 0.998 & 0.997 & 0.992 &  & 0.812 & 0.698 & 0.609\\
 &  &  & \scriptsize{$(0.995,1.000)$} & \scriptsize{$(0.994,1.000)$} & \scriptsize{$(0.986,0.998)$} &  & \scriptsize{$(0.788,0.836)$} & \scriptsize{$(0.670,0.726)$} & \scriptsize{$(0.579,0.639)$}\\
\hline
\end{tabular}
\end{table}

\section{Discussion and extensions} \label{Discussion&extensions}

This paper has proposed a test for a parametric specification of the variance function in the nonparametric regression model \eqref{model}. It has some advantages over some previously existing tests. To apply the test, it is assumed that the available data consist of independent observations from  $(X,Y)^T$. Several extensions are possible. Specifically, the procedure could be extended to mo\-dels with more than one covariate and to dependent data. Next we briefly sketch the case of dependent data and leave the case of multivariate covariate for future research.

Let $(X_j,Y_j)^T$, $j=0, \pm 1, \pm 2, \ldots$, be a bivariate strictly stationary  discrete time process  satisfying \eqref{model}, that is, $Y_j=m(X_j)+\sigma(X_j)\varepsilon_j$. We also assume that  $\varepsilon_j$  is  independent of $X_j$ and  that $\{\varepsilon_j\}_{j \in \mathbb{Z}}$ are independent and identically distributed with mean 0 and variance 1. This general nonparametric framework includes typical time series models, where $X_j$
represents lagged variables of $Y_j$ (for instance $X_j =Y_{j-1}$). An example is the ARCH(1) model (see, for example, Fan and Yao, 2003, p. 143),
\[
Y_j=(\theta_1+\theta_2Y^2_{j-1})^{1/2}\varepsilon_j,
\]
for some constants $\theta_1, \theta_2 \geq 0$, $\theta_2<1$, where $\varepsilon_j$ has mean 0 and variance 1 and is independent
of $Y_{j-1}$ for all $j$. Motivated by this fact, and in contrast to the case of independent data, the assumption that the covariate has a compact support will be removed in this setting.

Let  $(X_1,Y_1)^T,  \ldots, (X_n,Y_n)^T$ be  generated from  model \eqref{model} with the above assumptions. As in Section \ref{theteststatistic}, $m$, $\sigma$ and $\theta$ are replaced by $\hat{m}$,  $\hat{\sigma}$ and $\hat{\theta}$ to calculate the ordinary and null residuals, with
$\hat{m}$,  $\hat{\sigma}$  as in  \eqref{m&s}. If the covariate takes values in an unbounded interval, then it is not possible to get convenient convergence rates for $\hat{m}$ and  $\hat{\sigma}$ to ${m}$ and  ${\sigma}$, respectively, in the whole interval (see Hansen, 2008). Because of this reason, we slightly modify the definition of the empirical characteristic functions associated with these residuals as follows:
\[
 \hat{\varphi}_{g}(t)=\frac{1}{n}\sum_{j=1}^n g(X_j)\exp (\i t \hat{\varepsilon}_{j}),\quad
 \hat{\varphi}_{0g}(t)=\frac{1}{n}\sum_{j=1}^n g(X_j)\exp (\i t \hat{\varepsilon}_{0j}),
 \]
where $g$ is a non-negative weight function with compact support $R_g$. A simple choice for $g$ is the indicator function of a compact interval where $X$ takes values with high probability. This choice can be seen as the natural extension of the independent case since in such context we took $g$ as  the indicator function on the interval $R$, the support of $X$. Nevertheless, the theory that will be developed is valid for any bounded function $g$ with compact support. With this consideration, the null hypothesis becomes
 \[
H_{0} : \sigma^2(x)= \sigma^2(x; \theta), \quad \mbox{for some }\theta \in \Theta \subset \mathbb{R}^p,  \quad \forall x \in R_g.
\]

As in the independent data setting, to measure the distance between the empirical characteristic functions $\hat{\varphi}_g$ and $\hat{\varphi}_{0g}$ we will employ an $L_2(w)$ norm, for some non-negative function $w$, that is,
 \[
 T_{n,g}=\|\hat{\varphi}_g-\hat{\varphi}_{0g} \|_w^2.
 \]

 In order to calculate the asymptotic null distribution  of $T_{n,g}$  it will be  also  assumed that the process $(X_j,Y_j)^T$, $j=0, \pm 1, \pm 2, \ldots$,  is
absolutely regular, also called  $\beta$-mixing, which means that (see, for example, Bradley, 2005)
\[
\beta_j=\sup_{k \in \mathbb{Z}}\beta(\mathcal{F}_{-\infty}^k, \mathcal{F}^{\infty}_{k+j})\to 0,
\]
as $j \to \infty$, where $\mathcal{F}_L^J$ stands for the $\sigma$-algebra generated by $\{(X_j,Y_j)^T, L\leq j \leq J\}$, $-\infty \leq L \leq J \leq \infty$, and for two $\sigma$-algebras $\mathcal{A}$ and $\mathcal{B}$,
\[
\beta(\mathcal{A},\mathcal{B})= \sup \frac{1}{2} \sum_{j=1}^J \sum_{k=1}^K |P(A_j \cap B_k)-P(A_j)P(B_k)|,
\]
with the supremum  taken over all pairs of (finite) partitions $\{A_1, \ldots ,A_J \}$ and  $\{B_1, \ldots ,B_K \}$
such that $A_j\in \mathcal{A}$ for each $j$ and $B_k\in \mathcal{B}$ for each $k$.

The following  regularity assumption will be used to prove the results for the current case.

\newpage

\begin{description}

\item[{\bf Assumption C:}]

\item[\qquad C.1]  The mixing coefficients satisfy  $\beta_j=O(j^{-b})$, for some $b>2$.
 
\item[\qquad C.2]  (i) $X_j$ is absolutely continuous with pdf $f$.  (ii) $g$ has compact support $R_g$, $g(x)> 0$, for all $x\in R_g$, $\sup_x g(x)<\infty$. (iii) $f$, $g$, $m$ and $\sigma$ are two times continuously differentiable on $R_g$. (iv) $\inf_{x\in R_g} f(x)>0$ and $\inf_{x\in R_g} \sigma(x)>0$. (v) $E(\varepsilon^4)<\infty$.

\item[\qquad C.3] (i) $E(|Y_j|^s)<\infty$ and $\sup_{x \in R_g}  E  (|Y_j|^s \, | \, X_j=x)<\infty$, for some $s>2+2/(b-2)$. (ii) There exists some  $j^*$ such that for al $j \geq j^*$
\[
\sup_{x_0,x_j \in R_g}E(|Y_0 Y_j| \, | \, X_0=x_0, X_j=x_j)f_j(x_0,x_j)<\infty,
\]
where $f_j(x_0,x_j)$ denotes the joint density of $(X_0,X_j)$.

\item[\qquad C.4] $\ln n/hn^{\gamma}\to 0$ for $\gamma=\min\{0.5,(b-2-(1+b)/(s-1))/(b+2-(1+b)/(s-1))\}$.

\item[\qquad C.5] The estimator $\hat{\theta}$ satisfies
\[
\sqrt{n}(\hat{\theta}-{\theta}_0)=\frac{1}{\sqrt{n}}\sum_{j=1}^nl(\varepsilon_j, X_j; \theta_0)+o_P(1),
\]
with $E\{l(\varepsilon_j, X_j; \theta_0) \, | \, X_j\}=0$ and $E\{l(\varepsilon_j, X_j; \theta_0)^{2+\varrho}\}<\infty$, for some $\varrho>0$.

\item[\qquad C.6] (i) $\sigma^2(x;\theta)$ is    twice continuously differentiable with respect to $\theta$ and the derivatives are continuous in $(x,\theta)$, for all $x$ and $\theta $. (ii) $\inf_{x\in R_g, \, \theta\in \Theta_0} \sigma(x;\theta)>0$, for some $\Theta_0 \subseteq \Theta$ such that $\theta_0 \in int\Theta_0$.
\end{description}

As in Subsection
\ref{AsymptoticNullDistribution}, we first give a result that provides an asymptotic approximation for $\sqrt{n}\{\hat{\varphi}_g(t)-\hat{\varphi}_{0g}(t)\}$ when the null hypothesis is true. Let
\[\mu_g=E\{g(X)\dot \sigma^2(X;\theta_0)/\sigma^2(X;\theta_0)\}.\]

\begin{theorem} \label{null-expansion-dep}
Suppose that Assumptions A.2--A.3 and C hold. If $H_0$ is true, then
\begin{equation}   \label{expansion-th11}
\sqrt{n}\{\hat{\varphi}_g(t)-\hat{\varphi}_{0g}(t)\}=-\frac{t}{2} \varphi'(t)\frac{1}{\sqrt{n}}\sum_{j=1}^n\{(\varepsilon^2_j-1)g(X_j)-\mu_g^Tl(\varepsilon_j, X_j; \theta_0) \}+tR_1(t)+t^2R_2(t),
\end{equation}
with $\sup_t |R_k(t)|=o_P(1)$, $k=1,2$.
\end{theorem}

Now, by applying the central limit theorem for $\alpha$-mixing processes we get the following.

 \begin{corollary} \label{null-distribution-dep}
Suppose that the assumptions in Theorem \ref{null-expansion-dep} hold and that the weight function satisfies \eqref{condicionw}, then
\[nT_{n,g}  \stackrel{\mathcal{L}}{\longrightarrow}  \zeta^2Z^2,\]
where $Z$ is a standard normal variate, $\zeta^2=\zeta^2_1 \zeta^2_2 $, $\zeta^2_1=\frac{1}{4} \| t \varphi'(t) \|_w^2$  and $\zeta^2_2=\sum_{j=1}^{\infty}E[ \{(\varepsilon^2_1-1)g(X_1)-\mu_g^T l(\varepsilon_1, X_1; \theta_0) \}
\{(\varepsilon^2_j-1)g(X_j)-\mu_g^T l(\varepsilon_j, X_j; \theta_0) \}]$.
\end{corollary}

Observe that Corollary \ref{null-distribution} is a special case of Corollary \ref{null-distribution-dep} for independent data, $X$ taking values in a compact set $R$, $g$ being the indicator function in $R$ and $l(\varepsilon_j, X_j; \theta_0)=\Omega^{-1}(\varepsilon_j^2-1)\sigma^2(X_j;\theta_0)\dot \sigma^2(X_j;\theta_0)$
as in (\ref{thetanull}).

As in the independent data case,  the  asymptotic null distribution of $nT_{n,g}$  depends on unknown quantities. Similar strategies to those discussed in Section \ref{sect.prac} could be also applied in this setting.

\begin{remark} In the above development we have assumed that the function $g$ does not vary with $n$ but, as noticed by an anonymous referee, it can be chosen depending on $n$. Specifically, we could consider
\begin{equation}   \label{gdependn}
 \hat{\varphi}_{g_n}(t)=\frac{1}{n}\sum_{j=1}^n g_n(X_j)\exp (\i t \hat{\varepsilon}_{j}),\quad
 \hat{\varphi}_{0g_n}(t)=\frac{1}{n}\sum_{j=1}^n g_n(X_j)\exp (\i t \hat{\varepsilon}_{0j}),
 \end{equation}
$g_n$ being a  non-negative weight function with  support $R_{g_n}$, so that $\sup_{x \in \mathbb{R}}g_n(x) \leq M$, for some positive $M$, $\forall n\in \mathbb{N}$. Suppose that $R_{g_n} \subseteq R_g$, that $\sup_{x \in \mathbb{R}}|g_n(x)-g(x)| \to 0$ and that the assumptions in Theorem \ref{null-expansion-dep} hold, then the expansion for
$\sqrt{n}\{\hat{\varphi}_g(t)-\hat{\varphi}_{0g}(t)\}$ in \eqref{expansion-th11} is also true for 
$\sqrt{n}\{\hat{\varphi}_{g_n}(t)-\hat{\varphi}_{0g_n}(t)\}$. 

A key assumption in Theorem \ref{null-expansion-dep} is that  $R_g$ is compact. This assumption can be dropped. Nevertheless, in order to get similar convergence rates to those in \eqref{esto2} required in our proofs, we must adopt the weighting scheme in \eqref{gdependn}  and strengthen  Assumption C. Compare, for example, with the somehow related developments in Neumeyer and Selk (2013) and Selk and Neumeyer (2013) when dealing with a weighted empirical distribution function of the residuals.
\end{remark}

\appendix

\section{Proofs}

\subsection{Sketch of the proofs of results in Section \ref{Main results}}

Observe  that under Assumptions A.1, A.2 and A.3  for $(X_1,Y_1), \ldots, (X_n,Y_n)$  independent from \eqref{model}, (see, for example, Hansen, 2008), 
\begin{equation} \label{esto}
\begin{array}{lll}
\displaystyle
\sup_{x\in R}|\hat{m}(x)-m(x)|=o_P(n^{-1/4}),\\
\displaystyle
\sup_{x\in R}|\hat{\sigma}(x)-\sigma(x)|=o_P(n^{-1/4}). \\
\end{array}
\end{equation}
\noindent {\bf Proof of (\ref{thetanullprobability})} \hspace{2pt}
Let $S_n(\theta)$ and $S(\theta)$ be as defined in \eqref{ese_ene} and \eqref{ese}, respectively. From   \eqref{esto} and the SLLN, it follows that
$$S_n(\theta)=S(\theta)+o_P(1), \quad \forall \theta \in \Theta.$$
Next we prove that the above convergence holds uniformly in $\theta$. Routine calculations show that  the  derivative
\[\frac{\partial}{\partial \theta}\left\{S_n(\theta)-S(\theta)\right\},
\]
is bounded (in probability) $\forall \theta \in \Theta_0$, which implies that the family $\left\{S_n(\theta)-S(\theta), \, \theta \in \Theta_0\right\}$ is equicontinuous. By the Lemma in Yuan (1997), it follows that
\[\inf_{\|\theta-\theta_0\|>\delta} S_n(\theta)-S_n(\theta_0)=\inf_{\|\theta-\theta_0\|>\delta} S(\theta)-S(\theta_0)+o_p(1).\]
Now the result follows from Assumption B.1 and Lemma 1 in Wu (1981).
$\Box$

\bigskip

\noindent {\bf Proof of (\ref{thetanull})} \hspace{2pt}
Under the assumptions made, routine calculations show that under $H_0$,
\begin{eqnarray}
\sqrt{n} \frac{\partial}{\partial \theta} S_n(\theta_0) & = & -\frac{2}{\sqrt{n}}\sum_{j=1}^n(\varepsilon^2_j-1)\sigma^2(X_j; \theta_0)\dot \sigma^2(X_j; \theta_0) +o_P(1), \label{auxnew1}\\
\frac{\partial^2}{\partial \theta \partial \theta^{T}} S_n(\theta_0) & = & 2 \Omega+o_P(1). \label{auxnew2}
\end{eqnarray}
By Taylor expansion,
\begin{equation} \label{auxnew3}
\frac{\partial}{\partial \theta} S_n(\hat{\theta})-\frac{\partial}{\partial \theta} S_n(\theta_0)=\frac{\partial^2}{\partial \theta \partial \theta^{T}} S_n(\theta_0)(\hat{\theta}-\theta_0)+o(\|\hat{\theta}-\theta_0\|).
\end{equation}
Taking into account that $\frac{\partial}{\partial \theta} S_n(\hat{\theta})=0$, the result follows from \eqref{auxnew1}--\eqref{auxnew3}.
$\Box$

\bigskip

\noindent {\bf Proof of Theorem \ref{null-expansion}} \hspace{2pt}
From Lemma 10 (i) in Pardo-Fern\'andez et al.
 (2015b) it follows that
\[
\begin{array}{rcl}
\displaystyle \sqrt{n}\hat{\varphi}(t) & = & \displaystyle \sqrt{n}\tilde{\varphi}(t)+\i \frac{t}{\sqrt{n}}\sum_{j=1}^n \exp(\i t\varepsilon_j)\frac{m(X_j)-\hat{m}(X_j)}{\sigma(X_j)}\\
 & & \displaystyle +
\i \frac{t}{\sqrt{n}}\sum_{j=1}^n \exp(\i t\varepsilon_j)\varepsilon_j\frac{\sigma(X_j)-\hat{\sigma}(X_j)}{\sigma(X_j)}+tR_{11}(t)+t^2R_{12}(t),
\end{array}
\]
with
\begin{equation} \label{aux00}
\tilde{\varphi}(t)=\frac{1}{n}\sum_{j=1}^n\exp(\i t \varepsilon_j)
\end{equation}
and $\sup_t |R_{1k}(t)|=o_P(1)$, $k=1,2$. As for  $\hat{\varphi}_0(t)$, since
\[
\begin{array}{rcl}
\displaystyle
\hat{\varepsilon}_{0j}-\varepsilon_j & = &
\displaystyle
\frac{m(X_j)-\hat{m}(X_j)}{\sigma(X_j)}+
\frac{\{m(X_j)-\hat{m}(X_j) \} \{ \sigma(X_j)-\sigma(X_j;\hat{\theta}) \}}{\sigma(X_j)\sigma(X_j;\hat{\theta})}\\
 & & \displaystyle
+ \frac{ \sigma(X_j)-\sigma(X_j;\hat{\theta}) }{\sigma(X_j)}\varepsilon_j+
 \frac{ \{ \sigma(X_j)-\sigma(X_j;\hat{\theta}) \}^2}{\sigma(X_j)\sigma(X_j;\hat{\theta})}\varepsilon_j,
\end{array}
\]
we have that
\[
\begin{array}{rcl}
\displaystyle \sqrt{n}\hat{\varphi}_0(t) & = & \displaystyle \sqrt{n}\tilde{\varphi}(t)+\i \frac{t}{\sqrt{n}}\sum_{j=1}^n \exp(\i t\varepsilon_j)\frac{m(X_j)-\hat{m}(X_j)}{\sigma(X_j)}\\
 & & \displaystyle +
\i \frac{t}{\sqrt{n}}\sum_{j=1}^n \exp(\i t\varepsilon_j)\varepsilon_j\frac{\sigma(X_j)-{\sigma}(X_j;\hat{\theta})}{\sigma(X_j)}+tR_{13}(t)+t^2R_{14}(t),
\end{array}
\]
with $\sup_t |R_{1k}(t)|=o_P(1)$, $k=3,4$. From Lemma 11 in Pardo-Fern\'andez et al.
 (2015b)
\[
\i \frac{t}{\sqrt{n}}\sum_{j=1}^n \exp(\i t\varepsilon_j)\varepsilon_j\frac{\sigma(X_j)-\hat{\sigma}(X_j)}{\sigma(X_j)}=-\frac{t}{2}\varphi'(t)\frac{1}{\sqrt{n}}\sum_{j=1}^n
(\varepsilon_j^2-1)+R_{15}(t)
\]
with $\|R_{15}\|_w=o_P(1)$. By Taylor expansion and \eqref{thetanull},
\[
\i \frac{t}{\sqrt{n}}\sum_{j=1}^n \exp(\i t\varepsilon_j)\varepsilon_j\frac{\sigma(X_j)-{\sigma}(X_j;\hat{\theta})}{\sigma(X_j)}=V(t)+R_{16}(t),
\]
with $\|R_{16}\|_w=o_P(1)$ and
\[
V(t)=\frac{1}{2}\frac{\i t}{n\sqrt{n}} \sum_{j,k=1}^n \exp(\i t\varepsilon_j)\varepsilon_j\frac{\dot \sigma^2(X_j;\theta_0)^T}{\sigma^2(X_j;\theta_0)} \Omega^{-1}
\dot \sigma^2(X_k;\theta_0)\sigma^2(X_k;\theta_0)(\varepsilon_k^2-1).
\]
Routine calculations show that
\[
V(t)=-\frac{t}{2}\varphi'(t)\frac{1}{\sqrt{n}}\mu^T\Omega^{-1}\sum_{j=1}^n\dot \sigma^2(X_j;\theta)\sigma^2(X_j;\theta)(\varepsilon_j^2-1)+R_{17}(t)
\]
with $\|R_{17}\|_w=o_P(1)$. Therefore the result follows. $\Box$

\bigskip

\noindent {\bf Proof of Theorem \ref{limite}} \hspace{2pt}
From Lemma 10 (i) in Pardo-Fern\'andez et al.
 (2015b) it follows that
\begin{equation} \label{aux1}
\hat{\varphi}(t)=\tilde{\varphi}(t)+tR(t),
\end{equation}
with $\tilde{\varphi}(t)$ as defined in \eqref{aux00} and
\begin{equation} \label{aux2}
 \quad \sup_t |R(t)|=o_P(1).
\end{equation}
As for  $\hat{\varphi}_0(t)$, since
\[
\hat{\varepsilon}_{0j}-\varepsilon_{0j}  = \frac{m(X_j)-\hat{m}(X_j)}{\sigma(X_j;\hat{\theta})}+\varepsilon_j\frac{\sigma(X_j)}{\sigma(X_j;\hat{\theta})\sigma(X_j; \theta_0)}
\left\{\sigma(X_j;\theta_0)-\sigma(X_j;\hat{\theta})\right\},
\]
\[
\left | \frac{1}{n}\sum_{j=1}^n\frac{m(X_j)-\hat{m}(X_j)}{\sigma(X_j;\hat{\theta})} \right|  \leq  \frac{1}{\displaystyle \inf_{x\in R, \, \theta\in \Theta_0} \sigma(x;\theta)}\sup_{x\in R}|\hat{m}(x)-m(x)|=o_P(1),
\]
and
\[
\left | \frac{1}{n}\sum_{j=1}^n \varepsilon_j\frac{\sigma(X_j)}{\sigma(X_j;\hat{\theta})\sigma(X_j; \theta_0)}
\left\{\sigma(X_j;\theta_0)-\sigma(X_j;\hat{\theta})\right\} \right| \] \[\leq \left(\frac{1}{n}\sum_{j=1}^n \varepsilon_j^2\right)^{1/2}
\frac{\displaystyle \sup_{x\in R} \sigma(X_j)}{\displaystyle \inf_{x\in R, \, \theta\in \Theta_0} \sigma^3(x;\theta)}
\left(\frac{1}{n}\sum_{j=1}^n \left\{\sigma^2(X_j;\theta_0)-\sigma^2(X_j;\hat{\theta})\right\}^2\right)^{1/2}=o_P(1),
\]
by Taylor expansion we get
\begin{equation} \label{aux3}
\hat{\varphi}_0(t)=\tilde{\varphi}_0(t)+tR_0(t),
\end{equation}
with
\begin{equation} \label{aux4}
\tilde{\varphi}_0(t)=\frac{1}{n}\sum_{j=1}^n\exp(\i t \varepsilon_{0j}), \quad \sup_t |R_0(t)|=o_P(1).
\end{equation}
The result follows from \eqref{aux1}--\eqref{aux4}, by taking into account that $\|\tilde{\varphi}-{\varphi}\|_w=o_P(1)$ and $\|\tilde{\varphi}_0-{\varphi}_0\|_w=o_P(1)$.
$\Box$

\bigskip

\noindent {\bf Proof of Theorem \ref{cont-expansion}} \hspace{2pt}  Under $H_{1n}$,
\[ \sqrt{n}(\hat{\theta}-{\theta}_0)  =
\displaystyle \Omega^{-1}\frac{1}{\sqrt{n}}\sum_{j=1}^n(\varepsilon_j^2-1)\sigma^2(X_j;\theta_0)\dot \sigma^2(X_j;\theta_0)
 \displaystyle +\Omega^{-1}E\{r(X)\dot \sigma^2(X;\theta_0)\}+o_P(1).
\]
By applying the results in Yuan (1997), we get that \eqref{esto} also hold under $H_{1n}$. Now, the result follows by proceeding similarly to the proof of Theorem \ref{null-expansion}. $\Box$

\subsection{Sketch of the proofs of results in Section \ref{Discussion&extensions}}
Under the assumptions A.2, A.3, C.1--C.4 and C.6  (see, for example, Hansen 2008),
\begin{equation} \label{esto2}
\begin{array}{lll}
\displaystyle
\sup_{x\in R_g}|\hat{m}(x)-m(x)|=o_P(n^{-1/4}),\\
\displaystyle
\sup_{x\in R_g}|\hat{\sigma}(x)-\sigma(x)|=o_P(n^{-1/4}). \\
\end{array}
\end{equation}

\bigskip

\noindent {\bf Proof of Theorem \ref{null-expansion-dep}} \hspace{2pt}
Proceeding as in the proof of Theorem \ref{null-expansion}, we obtain
\[
\sqrt{n} \left\{\hat{\varphi}_g(t)-\hat{\varphi}_{0g}(t)\right\} = V_1(t)+V_2(t)+tR_1(t)+t^2R_2(t),
\]
with   $\sup_t |R_{k}(t)|=o_P(1)$, $k=1,2$,
\begin{eqnarray*}
V_1(t) & = & \i \frac{t}{\sqrt{n}}\sum_{j=1}^n \exp(\i t\varepsilon_j)\varepsilon_jg(X_j)\frac{\sigma(X_j)-\hat{\sigma}(X_j)}{\sigma(X_j)},\\
V_2(t) & = & \i \frac{t}{\sqrt{n}}\sum_{j=1}^n \exp(\i t\varepsilon_j)\varepsilon_jg(X_j)\frac{\sigma(X_j)-\sigma(X_j;\hat{\theta})}{\sigma(X_j)}.
\end{eqnarray*}
From \eqref{esto2},
\[
\sup_{x \in R_g}\left|\hat{\sigma}(x)-\sigma(x)-\frac{1}{2nf(x)\sigma(x)}\sum_{j=1}^{n}
K_h(X_{j}-x)\left[\left\{Y_{j}-m(x)\right\}^2-\sigma^2(x)\right]\right|=o_p(n^{-1/2}),
\]
where $K_h(\cdot)=\frac{1}{h}K(\frac{\cdot}{h})$. By using this identity we get that
\[V_1(t)=V_3(t)+tR_3(t),\]
with $\sup_t |R_{3}(t)|=o_P(1)$ and
\[
V_3(t)=-\i \frac{t}{2n\sqrt{n}}\sum_{j,k=1}^n \exp(\i t\varepsilon_j)\varepsilon_jg(X_j)\frac{1}{\sigma^2(X_j)f(X_j)}K_h(X_{j}-X_k)\left[\left\{Y_k-m(X_j)\right\}^2-\sigma^2(X_j)\right].
\]
By applying Hoeffding decomposition and Lemma 2 in Yoshihara (1976) we get that
\[
V_3(t)=-\frac{t}{2} \varphi'(t)\frac{1}{\sqrt{n}}\sum_{j=1}^ng(X_j)(\varepsilon^2_j-1)+tR_4(t),
\]
with $\sup_t |R_{4}(t)|=o_P(1)$.

By Taylor expansion and following similar steps to those given for $V_1(t)$, we obtain
\[
V_2(t)=-\frac{t}{2} \varphi'(t)\frac{1}{\sqrt{n}}\sum_{j=1}^n\mu_g^T   l(\varepsilon_j, X_j; \theta_0)+tR_5(t),
\]
with $\sup_t |R_{5}(t)|=o_P(1)$. Putting together all above facts the result follows.
$\Box$

\section*{Acknowledgements}

The authors thank the anonymous referees for their valuable time and careful comments, which improved the presentation of this paper. The authors acknowledge financial support from grants MTM2014-55966-P and MTM2017-89422-P, funded by the Spanish Ministerio de Econom\'{\i}a, Industria y Competitividad, the Agencia Estatal de Investigaci\'{o}n and the European Regional Development Fund. J.C. Pardo-Fern\'{a}ndez also acknowledges funding from Banco Santander and Complutense University of Madrid (project PR26/16-5B-1). M.D. Jim\'{e}nez-Gamero also acknowledges support from CRoNoS COST Action IC1408.

\section*{References}

\bib
Alba-Fern\'andez V, Jim\'enez-Gamero MD,  Mu\~{n}oz-Garc\'{\i}a J (2008)
A  test for the two-sample problem based on empirical characteristic functions. 
\emph{Comput. Stat. Data. Anal.} 52, 3730--3748.

\bib
Bradley, RC (2005)
Basic properties of strong mixing conditions. A survey and some open questions.
\emph{Probab. Surv.} 2, 107--144.

\bib
Dette H, Hetzler B (2009a)
A simple test for the parametric form of the variance function in nonparametric regression.
\emph{Ann. Inst. Statist. Math.} 61, 861--886.

\bib
Dette H, Hetzler B (2009b)
Khmaladze transformation of integrated variance processes with applications to goodness-of-fit testing.
\emph{Math. Methods Statist.} 18,  97--116.

\bib
Dette H, Marchlewski M (2010)
A robust test for homoscedasticity in nonparametric regression.
\emph{J. Nonparametr. Stat.} 22, 723--736.

\bib
Dette H, Neumeyer N, Van Keilegom I (2007)
A new test for the parametric form of the variance function in non-parametric regression.
\emph{J. Roy. Statist. Soc. Ser. B} 69, 903--917.

\bib
Fan J,  Gijbels I (1996)
\emph{Local Polynomial Modelling and Its Applications.}
Champan \& Hall, London.

\bib
Fan J,  Yao Q (2003)
\emph{Nonlinear time series. Nonparametric and parametric methods.}
Springer, New York.

\bib
Feller W (1971)
\emph{An Introduction to Probability Theory and its Applications, Vol 2.}
Wiley.

\bib
Hansen BE (2008)
Uniform convergence rates for kernel estimation with dependent data.
\emph{Econometric Theory} 24, 726--748.

\bib
Hu\v{s}kov\'a M,  Meintanis SG (2009) Goodness-of-fit tests for parametric regression models based on empirical characteristic functions. \emph{Kybernetika}  45, 960--971.

\bib
Hu\v{s}kov\'a M,  Meintanis SG (2010) Tests for the error distribution in nonparametric possibly heteroscedastic regression models. \emph{TEST}  19, 92--112.

\bib
Koul HL, Song W (2010)
Conditional variance model checking.
\emph{J. Statist. Plann. Inference} 140, 1056--1072.

\bib
Liero H (2003)
Testing homoscedasticity in nonparametric regression.
\emph{J. Nonparametr. Stat.} 15, 31--51.

\bib
Neumeyer N, Selk  L (2013) 
A note on non-parametric testing for Gaussian innovations in AR-ARCH models. 
\emph{J. Time Series Anal.} 34, 362--367. 

\bib
Neumeyer N, Van Keilegom I (2017)
Bootstrap of residual processes in regression: to smooth or not to smooth?
(available at arXiv:1712.02685v1).

\bib
Pardo-Fern\'andez JC, Jim\'enez-Gamero MD, El Ghouch A (2015a)
A nonparametric ANOVA-type test for regression curves based on characteristic functions.
\emph{Scand. J. Stat.} 42, 197--213.

\bib
Pardo-Fern\'andez JC, Jim\'enez-Gamero MD, El Ghouch A (2015b)
Tests for the equality of conditional variance functions in nonparametric regression.
\emph{Electron. J. Stat.} 9, 1826--1851.

\bib
Samarakoon N, Song W (2011)
Minimum distance conditional variance function checking in heteroscedastic regression models.
\emph{J. Multivariate Anal.} 102, 579--600.

\bib
Samarakoon N, Song W (2012)
Empirical smoothing lack-of-fit tests for variance function.
\emph{ J. Statist. Plann. Inference} 142, 1128--1140.

\bib
Selk L, Neumeyer N (2013)  
Testing for a change of the innovation distribution in nonparametric autoregression: the sequential empirical process approach. 
\emph{Scand. J. Stat.} 40, 770--788.

\bib
Wang L, Zhou X-H (2007)
Assessing the adequacy of variance function in heteroscedastic regression models.
\emph{Biometrics} 63, 1218--1225.

\bib
Wu CF (1981) Asymptotic theory of nonlinear least squares estimation.
\emph{Ann. Statist.} 9, 501--513.

\bib
Yoshihara KI (1976)
Limiting behavior of U-statistics for stationary, absolutely regular processes.
\emph{Z. Wahrsch. Verw. Gebiete} 35, 237--252.

\bib
Yuan KH (1997)
A theorem on uniform convergence of stochastic functions with applications.
\emph{J. Multivariate Anal.} 62, 100--109.

\end{document}